\documentclass[fp]{jpsj3}
\usepackage{amsthm,amsmath}
\usepackage{latexsym}
\usepackage{txfonts}
\usepackage{graphicx}
\usepackage{bm}
\usepackage{array}
\usepackage{color}
\usepackage{ulem}

\begin{document}

\title{DC-Current Induced Domain Wall in a Chiral $p$-Wave Superconductor}

\author{Thibaut Jonckheere$^{1}$ and Takeo Kato$^{2}$}
\inst{${^1}$Aix Marseille Univ, Universit\'e de Toulon, CNRS, CPT, Marseille, France\\
${^2}$Institute for Solid State Physics, University of Tokyo, Kashiwa, 277-8581, Japan.} 

\abst{We study theoretically the impact of an applied DC-current on a mesoscopic chiral $p$-wave superconductor. 
Performing quasi-classical calculations on a two-dimensional system, with an external magnetic flux to generate a DC current, we show that the current can trigger a transition to a state with a domain wall between regions of different chiralities. 
The system shows an hysteretic behavior, as different domain wall configurations are possible for a given current. 
This domain wall creation mechanism can give new insights on recent experiments observing anomalous current variations in Sr${}_2$RuO${}_4$ junctions.}


\maketitle

\maketitle

\section{Introduction} 

Chiral superconductors have been attracting interest by their striking features originating from time-reversal symmetry breaking and finite angular momentum of Cooper pairs.\cite{Kallin16}
Sr${}_2$RuO${}_4$, which is thought to be a chiral triplet $p$-wave superconductor, is one of the best-studied systems.\cite{Maeno94,Mackenzie03,Maeno12}
The chiral superconducting state, which is topologically non-trivial, exhibits topologically-protected chiral edge channels at the surfaces,\cite{Matsumoto99,Furusaki01} and has potential application to quantum computing using Majorana fermions in vortices.\cite{Ivanov01,Nayak08,Beenakker13,DasSarma15}

A number of experiments have been performed to characterize the superconductivity in Sr${}_2$RuO${}_4$.
The triplet nature of Sr${}_2$RuO${}_4$ has been observed by the Knight shift measurement,\cite{Ishida98} while spontaneous time-reversal symmetry breaking has been detected by muon spin resonance\cite{Luke98} and the Kerr effect.\cite{Xia06}
Edge currents predicted for chiral $p$-wave pairing symmetry, however, have not been observed in scanning SQUID experiments,\cite{Kirtley07,Hicks10} which has stimulated several theoretical studies proposing other symmetries for boundary-induced\cite{Tada09} or bulk\cite{Raghu10,Huang14,Komendova14} superconductivity, or considering the multi-band effect on chiral $p$-wave superconductivity.\cite{Taylor12,Wysokinski12,Wysokinski13,Imai12,Imai13,Scaffidi15,Kawai17,Zhang17a,Zhang17b}

Josephson junction experiments play an important role to determine pairing symmetry.
SQUID experiments for Sr${}_2$RuO${}_4$ have observed anomalous dynamical shift of the diffraction patterns.\cite{Kidwingira06,Saitoh15}
In addition, anomalous current-driven switching has been reported in $I$-$V$ characteristics of Josephson junctions.\cite{Kambara08,Kambara10,Anwar13,Anwar17}
These anomalies have been attributed to domain wall motion, which is compatible with chiral pairing symmetry, and indicate a pathway to novel devices which control the inertial degree of freedom in superconducting states. While theoretical studies of domain wall formation in Sr${}_2$RuO${}_4$ based on energetic arguments do exist,\cite{Vakaryuk}
details on how domain wall motion is driven by applied currents have not been clarified so far.

\begin{figure}[t]
\vspace{0.1cm}
\centering \includegraphics[width=7.5cm]{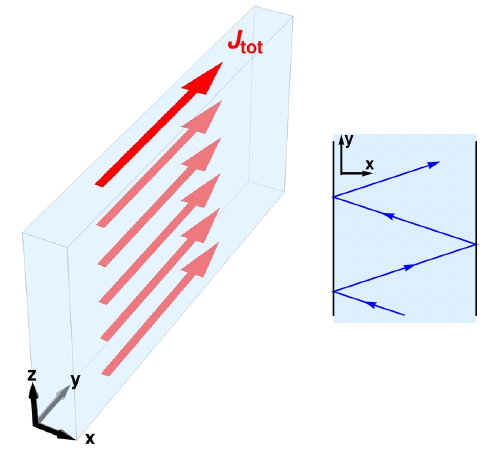}
\caption{(Color online) Left: a schematic view of the system. 
A slab of $p$-wave superconductor, with a width $L$ along the $x$ direction, and very large dimensions along the $y$ and $z$ direction.
 By applying an external vector potential, we induce a total current $J_{tot}$ along the $y$ direction, and the behavior of the system is essentially two-dimensional. 
Right: a typical classical trajectory along the 2d $x$-$y$ system, with specular reflection on both sides, which is used in the semi-classical calculations.}

\label{fig:system}
\end{figure}

The goal of this work is to study the impact of a DC-current bias on a $p$-wave superconductor with a finite width (see the left panel of Fig.~\ref{fig:system}).
We show that above a given threshold, a DC-current can create a domain wall separating two regions of different chiralities.
Because of this domain wall creation, the system shows an hysteretic behavior: for a given DC-current, it can be in different states depending on its history.
To keep the calculations as simple as possible, we consider the system at equilibrium, with the DC current created by the application of an external vector potential.
The results we obtain are relevant to understand experimental results of $I$-$V$ curves measured in current-biased weak link experiments.\cite{Kambara08,Kambara10,Anwar13} 
Our goal is to show that in addition to the domain wall motion, domain wall creation or destruction is also an essential mechanism of these systems when a DC-current is applied.

We employ the quasi-classical theory to evaluate the spatial profile of the pair potentials and of
the current.\cite{Schopohl98,Kopnin01}
By assuming that the superconducting gap is much smaller than the Fermi energy, this method can calculate properties of superconductors with arbitrary pairing symmetries, and has been applied to various phenomena in chiral $p$-wave superconductors, such as spontaneous chiral edge channels and domain walls\cite{Matsumoto99,Furusaki01}, local densities of states in mesoscale superconducting islands,\cite{Nagai12} and paramagnetism.\cite{Suzuki14}
In this paper, we follow the algorithm of Ref.~\citen{Nagai12}.

This paper is organized as follows. In Sect.~\ref{sec:model}, we detail the model (a $p$-wave chiral superconductor strip threaded by an external flux), and give the essential steps of the quasi-classical theory used to make the calculations. 
Sect.~\ref{sec:results} is devoted to the exposition and to the discussion of the results. 
In Sect.~\ref{sec:WeakScreening}, we show the results that we obtain for the pair potential components and the current density in the weak screening case. 
These results serve as a model to understand the results for the strong screening case, given in Sect.~\ref{sec:StrongScreening}, which are more complex but also directly relevant for experiments. 
In Sect.~\ref{sec:Discussion} that we discuss their relevance to existing experimental works.
Finally, Sect.~\ref{sec:conclusion} contains the conclusions and the perspectives of this work, and Appendix~\ref{appendix} contains details about the numerical calculations.

\section{Method}
\label{sec:model}

\subsection{Model}

We consider a $p$-wave superconductor in the shape of  slab with a finite width $L$ (along which we place the $x$ axis), and with very large
dimensions along the $y$ and $z$ directions (much larger than any significant length in the system), see Fig.~\ref{fig:system}. As we will study the impact of an applied current along the $y$ axis, the system is invariant by translation along the $z$ direction, and we can simply 
perform 2d calculations in the $x$-$y$ plane.

To apply the current, we consider that  on a large scale the system forms a loop (with a radius $R \to \infty$ in the $y$-$z$ plane),
 and we apply through this loop an external magnetic flux $\Phi_{\rm ext}$, which can be described by a potential vector
  $\mathbf{A}_{\rm ext} = (0,A_{y,{\rm ext}},0)$ in the absence of the screening current, where $A_{y,{\rm ext}}=\Phi_{\rm ext}/R$ is a constant parameter.
This external magnetic flux induces a superconducting current along the $y$ direction, whose properties can be examined in a thermal equilibrium state.
Although the $I$-$V$ characteristics of real superconducting junctions (or superconducting weak links) are more complex because the system is driven into a non-equilibrium state for a finite voltage bias,
we expect that this simple setup is sufficient to explain the essence of the domain-wall transition due to an external current, as observed in the experimental $I$-$V$ characteristics.\cite{Kambara08,Kambara08,Anwar13}

\subsection{Quasiclassical Theory}

For calculation of the order parameters and the current, we employ the quasi-classical formalism, using the Eilenberger equation.
The principle of the methods can be found in the literature,\cite{Matsumoto99,Nagai12} we give here only the essential steps, stressing the original points of the present work. 

The quasi-classical Green function $\hat{g}(\mathbf{k}_F,\mathbf{r},\omega_n)$ describing the superconductor is defined as:
\begin{equation}
 \hat{g}(\mathbf{k}_F,\mathbf{r},\omega_n) = \left( \begin{array}{cc} g & f \\ -\tilde{f} & -g     \end{array} \right),
\end{equation} 
where the $\mathbf{k}_F$ dependence represents the symmetry of the order parameter, $\mathbf{r}$ is the coordinate dependence,
$\omega_n = \pi T (2n+1)$ is a Matsubara frequency, and the 2$\times$2 matrix structure is in the Nambu space. 
The quasi-classical Green function is normalized as:
\begin{equation}
\hat{g}^2 = \hat{1}.
\end{equation}
The Eilenberger equation is:
\begin{align}
& \! \! -i \mathbf{v}_F \! \cdot \! \nabla \hat{g}  =  [ \hat{h}, \hat{g} ], \\
& \hat{h} = \biggl( \begin{array} {cc} 
   i \omega_n \!- \!e \mathbf{v}_F \!\cdot\! \mathbf{A}(\mathbf{r})   & - \Delta(\mathbf{k_F},\mathbf{r}) \\
    \Delta(\mathbf{k}_F,\mathbf{r})^*   &  -i \omega_n \! + \!  e \mathbf{v}_F \!\cdot\! \mathbf{A}(\mathbf{r}) 
\end{array} \biggr),
\end{align}
where $\Delta(\mathbf{k}_F,\mathbf{r})$ is pairing potential, and $\mathbf{A}(\mathbf{r})$ is the vector potential.

A standard method to solve the Eilenberger equation, taking into account the Green function normalization, is to use the Ricatti parametrization.\cite{Schopohl98}
We write $\hat{g}$ in terms of the Ricatti amplitudes $a$ and $b$ (we now omit the $\mathbf{r}$,
$\mathbf{k}_F$  and $\omega_n$ dependence for brevity), with:
\begin{equation}
\hat{g} = \frac{-1}{1+ a b} \left( \begin{array}{cc}  1- a b & 2 i a \\ -2 i b & -(1- a b) \end{array} \right),
\label{eq:RicattiAmplitudes}
\end{equation}
The amplitudes $a$ and $b$ obey the Ricatti equations:
\begin{align}
 \mathbf{v}_F \cdot \nabla a & = -2 (\omega_n - i e \mathbf{v}_F \cdot \mathbf{A} ) a - \Delta^* a^2 + \Delta, \label{eq:Ricattia} \\
 \mathbf{v}_F \cdot \nabla b & = +2 (\omega_n - i e \mathbf{v}_F \cdot \mathbf{A} ) b + \Delta \, b^2 - \Delta^*, \label{eq:Ricattib}
\end{align}
As the derivate appears only in $\mathbf{v}_F\cdot \nabla$, the Ricatti equations can be written as 1d equations along the semi-classical trajectories (given by the direction of $\mathbf{v}_F$).

The right panel of Fig.~\ref{fig:system} shows a typical classical trajectory: a straight line which is specularly reflected on both sides of the sample. 
At the position of the reflection $\mathbf{r}_n$, the $y$ component of the wave-vector $\mathbf{k}_F$ is conserved, while the $x$ component is reversed at each reflection.
The Ricatti amplitude then obey the boundary conditions:
\begin{align}
 a(\mathbf{r}_n,\mathbf{k}_F|_{out} ) & =  a(\mathbf{r}_n,\mathbf{k}_F|_{in} ) \\
 b(\mathbf{r}_n,\mathbf{k}_F|_{out} ) & =  b(\mathbf{r}_n,\mathbf{k}_F|_{in} ) 
\end{align}
where $\mathbf{k}_F|_{in}$ and $\mathbf{k}_F|_{out}$ are the incoming and outgoing wave-vectors at the reflection point. 
Using the translational invariance along the $y$ direction, all quantities can be expressed as function of $x$ only.  At a given point $x$, a trajectory is fully parametrized by the angle $\theta_k$ between the wavevector at this point and the $x$ axis.

\subsection{Expression of the Physical Quantities}

The physical quantities are expressed in terms of the Green function component $f$, $\tilde{f}$ and $g$, which are given in terms of the Ricatti amplitudes in Eq.~(\ref{eq:RicattiAmplitudes}). 
For superconductors with $p$-wave chiral symmetry, the pair potential $\Delta(\mathbf{k}_F,\mathbf{r}) = \Delta_x\cos(\theta_k) + i\Delta_y \sin(\theta_k)$ is determined by a self-consistent gap equation
\begin{align}
& \left(  \begin{array}{c}  \Delta_x(x) \\ \Delta_y(x)  \end{array} \right) =
   T \, V_p \sum_{0<\omega_n<\omega_C} \int_0^{\pi} d\theta_k      \left(  \begin{array}{c}  2 \cos \theta_k \\ 2 \sin \theta_k  \end{array} \right) \nonumber
 \\  & \hspace{20mm} \times \left(f(\theta_k,\omega_n,x) + \tilde{f}^*(\theta_k,\omega_n,x) \right) ,
  \label{eq:Gaps_fg} \\
& \left(V_p \right)^{-1} = \log\left(\frac{T}{T_C}\right) + \sum_{0<n<\omega_C/(2 \pi T)} \frac{1}{n-1/2} ,
\end{align}
where $T_C$ is the superconducting transition temperature and $\omega_C$ is a cutoff energy.~\cite{Matsumoto99}
The current density along the $y$ direction is:
\begin{multline}
J_y(x) = -2 e v_F N(0) T \sum_{0<\omega_n<\omega_C} \int_0^{\pi/2} \!\! d \theta_k \; \sin \theta_k 
  \\ \times  \mathrm{Im}  \left(  g(\theta_k,\omega_n,x) + g(\pi - \theta_k,\omega_n,x) \right)
  \label{eq:Jy_fg}
\end{multline}
where $N(0)$ is the normal density of states per unit volume at the Fermi energy.

The magnetic field and the vector potential are obtained from integration of the current density using the Maxwell equation as
\begin{align}
\frac{dB_z}{dx}(x) &= - \mu J_y(x), \label{eq:Bz} \\
\frac{dA_y}{dx}(x) &= B_z(x), \label{eq:Ay}
\end{align}
under the boundary conditions, $B_z(0) = (\mu/2) J_{tot}$,  $B_z(L) = -(\mu/2) J_{tot}$ and $A_y(L/2) = A_{y,{\rm ext}}$,
where $J_{tot} = \int_0^L dx J_y(x)$ is the total current.
These boundary conditions are a due to the infinite slab geometry: a simple application of the Ampere law shows that the magnetic field outside the slab is constant, with a value
directly proportional to the total current in the sample, and opposite signs on both sides.
Using Eqs.~(\ref{eq:Gaps_fg})-(\ref{eq:Ay}), one can iteratively find a solution of the problem for a given value of the external vector potential $A_{y,{\rm ext}}$.
Starting from an initial guess for the values of the gaps $\Delta_x(x)$, $\Delta_y(x)$ and the current density $J_y(x)$, one computes the value of the vector potential $\mathbf{A} = (0,A_y(x))$, and then solve the Ricatti equations (Eqs.~(\ref{eq:Ricattia})-(\ref{eq:Ricattib})) to obtain new values for $\Delta_x(x)$, $\Delta_y(x)$ and $J_y(x)$.

\section{Results}
\label{sec:results}

In this paper, we always consider a wide superconductor strip ($L \gg \xi_0$), taking $L=25 \xi_0$, where $\xi_0=\hbar v_F/ (\pi \Delta_0)$ is the superconducting coherence length
 (with $\Delta_0 \equiv \Delta_{\rm bulk}(T=0)$).

We study two situations.
First, in Sect.~\ref{sec:WeakScreening}, the case of $\lambda_L=L=25\xi_0$ for which the effect of the screening current is not significant.
Here, $\lambda_L=(\mu n_s e^2/m)^{-1/2}$ is the London penetration depth, where $n_s$ is the density of condensed electrons, and $m$ is the electron mass.
Second, we consider the case of $\lambda_L=L/10=2.5\xi_0$ in Sect.~\ref{sec:StrongScreening}, for which the screening effect is so strong that the magnetic field is almost zero inside the strip except for the edge region.
We note that the latter situation corresponds to the experiments for Sr$_2$RuO$_4$,\cite{Kambara08,Kambara10} while the former situation may be realized if one tunes the temperature just below the transition temperature $T_C$.
All the results we show have been obtained using a temperature $T= 0.2 T_C$, and a cutoff frequency $\omega_C = 10 T_C$.
Finally, we discuss the relevance of our results to experiments for Sr$_2$RuO$_4$ in Sect.~\ref{sec:Discussion}.

\subsection{Weak Screening Case}
\label{sec:WeakScreening}

\begin{figure*}[tb]
\includegraphics[width=16.cm]{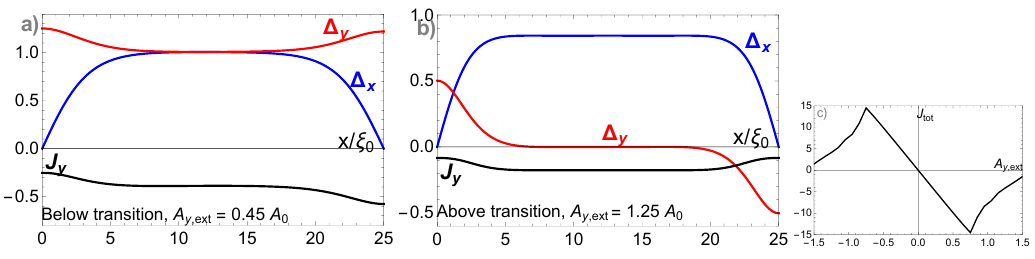}
\caption{(Color online) Results for the weak screening case ($\lambda_L=L=25\xi_0$).
The $p_x$- and $p_y$-components of the pair potential, $\Delta_x$ and $\Delta_y$ (normalized by the bulk zero-temperature gap $\Delta_0$) and the current density $J_y$ (normalized by $e v_F N(0) T_C$) as a function of $x$ for (a) below the transition, $A_{y,{\rm ext}}=0.45 A_0$ and (b) above the transition, $A_{y,{\rm ext}}=1.25 A_0$, where $A_0 = \Delta_0/(e v_F)$.
Below the transition, the pair potential components are similar to the zero applied flux case, the overall shift of the current is due to the applied flux, and the currents due to edge states are opposite on both sides of the sample. 
On the other hand, above the transition, the sign of $\Delta_y$ is inversed on the right side of the system, with a wide region near the middle where $\Delta_y$ is suppressed.
The current due to the edge states are now identical on both sides of the sample. 
(c) The total current $J_{tot}=\int_{0}^{L} dx J_y(x)$ as a function of  $A_{y,{\rm ext}}$. 
The total current is proportional to $A_{y,{\rm ext}}$, up to a transition value ($|A_{y,{\rm ext}}| \simeq 0.8 A_0$) beyond which it decreases in absolute value.}
\label{fig:lambda25}
\end{figure*}
 
The results for $\lambda_L=L=25\xi_0$ are shown in Fig.~\ref{fig:lambda25}.
Because the screening effect is weak, the external vector potential $A_{y,{\rm ext}}$ creates a superconducting current $J_y$ running in the $y$ direction which
is nearly uniform inside the whole width of the strip. This current is proportional to $A_{y,{\rm ext}}$, up to some value where a transition occurs. Above
the transition, the current decreases rapidly to 0.
This is visible
on the panel (c) of Fig.~\ref{fig:lambda25}, which shows the total current $J_{\rm tot}=\int_{0}^{L} dx J_y(x)$ as a function of $A_{y,{\rm ext}}$. One can see that the transition occurs for $|A_{y,{\rm ext}}| \simeq 0.8 A_0$ for our choice of parameters, where $A_0 = \Delta_0/(e v_F)$, with $\Delta_0 \equiv \Delta_{\rm bulk}(T=0)$.

The shape of the gap components (normalized
by $\Delta_0$) and of the current density (normalized by $e v_F N(0) T_C$) below the transition are shown on the panel (a) of Fig.~\ref{fig:lambda25}, for a value $A_{y,{\rm ext}} = 0.45 A_0$. One can see that the induced current is constant in the strip, except near the edges where the presence of the chiral edge currents, with opposite directions at the two edges, modifies $J_y$.\cite{Matsumoto99,Furusaki01}
For $A_{y,{\rm ext}}= 0.45 A_0$, the induced uniform current has a limited effect on the gap components $\Delta_x$ and $\Delta_y$, which are close to their values in the case without the applied flux ($A_{y,{\rm ext}}=0$); the system remains in a $p_x + i p_y$ state in the middle of the strip, while it has a $p_y$-like character at the two edges as $\Delta_x$ drops to zero there.

The profile of the gaps drastically changes if $A_{y,{\rm ext}}$ is larger than the transition threshold.
The superconducting gaps and the current density above the transition are shown on the panel (b) of Fig.~\ref{fig:lambda25}. 
One can see that the transition has had a dramatic effect on the $\Delta_y$ component, which is now negative near the right edge of the system, and close to zero around the center of the system. On the other hand,
the $\Delta_x$ gap component is not qualitatively changed from its value below the transition (there is the combination of a slight increase due to decrease of $\Delta_y$, because of the competing effect between two components,\cite{Matsumoto99}
and a global decrease due to large flux applied).
This means that the system is in a $p_x$ state in the middle of the strip, while below the transition it was in a $p_x + i p_y$ state. The change of sign of $\Delta_y$ is reflected on the sign of the edge currents: the edge currents have now the same signs on both edges of the system, while they had opposite signs below the transition.

For this low screening case, we found a unique stable solution for each value of $A_{y,{\rm ext}}$. We obtain the same results if the initial condition is the solution without external flux or if we perform a sweep and use as initial condition the solution obtained for a slightly different applied flux.
Therefore, there is no hysteresis in the flux-current relation for the weak screening case.

\subsection{Strong Screening Case}
\label{sec:StrongScreening}

\begin{figure*}[!htbp]
\includegraphics[width=16.cm]{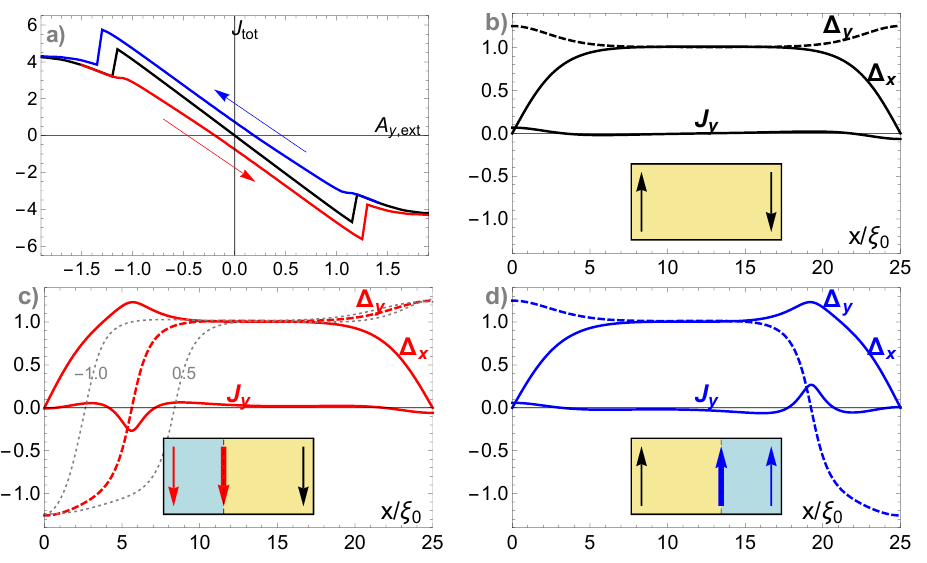} 
\caption{(Color online) Results for the strong screening case ($\lambda_L=L/10=2.5\xi_0$). 
(a) Total current as a function of the applied $A_{{\rm ext},y}$. 
The middle (black) curve corresponds to the solution obtained with the zero-field solution as initial state. 
The upper (blue) and lower (red) curves show the hysteresis when performing a sweep from a large negative $A_{y,{\rm ext}}$ value and from a large positive values of  $A_{y,{\rm ext}}$, respectively.
Other panels:  Plots of the values of $\Delta_x$, $\Delta_y$ and $J_y$ (normalized as in Fig.~\ref{fig:lambda25}) as a function of $x$ for the value of $A_{y,{\rm ext}}$ where the total current is 0.
(b) Non-sweep case, $A_{y,{\rm ext}}=0$. 
(c) The case of left to right sweep (in red, corresponding to the lower (red) curve in panel (a)) and (d) right to left sweep (in blue, corresponding to the upper (blue) curve in panel (a)), showing the existence of other stable solutions with zero total current: 
these solutions have a domain wall inside the system.
The insets on panels (b), (c) and (d) show a schematic picture of the edge currents directions (thin arrows) and of the domain wall current direction (thick arrows).
The two thin dotted gray curves of panel c) are plots of $\Delta_y$ for values of $A_{y,ext}=-1.0$ and $0.5$, which
show the motion of the domain wall as $A_{y,ext}$ is varied.}
\label{fig:lambda2p5}
\end{figure*}

The results for $\lambda_L=L/10=2.5\xi_0$ are shown in Fig.~\ref{fig:lambda2p5}. 
The panel (a) of Fig.~\ref{fig:lambda2p5} shows the total current $J_{\rm tot}=\int_{0}^{L} dx J_y(x)$ as a function of the applied $A_{y,{\rm ext}}$.
The middle (black) curve corresponds to the solution obtained with the zero-field solution as the initial state, indicating a behavior similar to the one of the low screening case: the total current is proportional to the applied flux $A_{y,{\rm ext}}$,  up to some value (here approximatively $|A_{y,{\rm ext}}| = 1.3 A_0$), where there is a transition.
We note that the critical current (the current at the threshold) is now much smaller than the one for the weak screening case, because the current is localized near the edges due to screening effect.

However, there exists other stable solutions in contrast to the weak screening case.
The lower (red) and the upper (blue) curves show the results obtained when performing a sweep from a large negative $A_{y,{\rm ext}}$ value and from a large positive values of $A_{y,{\rm ext}}$, respectively.
These two curves have been obtained by starting from an initial state beyond the transition, and then reducing $|A_{y,{\rm ext}}|$ gradually, using the solution at each step as the initial state for the next step. 
One can see that three stable solutions with different values of the total current $J_{\rm tot}$ are realized in a wide range of $A_{y,{\rm ext}}$.
The existence of three stable solutions means, for example, that it is possible to have the same total current $J_{\rm tot}=0$ with three different values of the applied flux $A_{y,{\rm ext}}$.
This is illustrated in Fig.~\ref{fig:lambda2p5}~(b)-(d).
Each of the graphs shows plots of $\Delta_x$, $\Delta_y$ and $J_y$ as a function of $x$ when the total current $J_{\rm tot}=0$. 
In Fig.~\ref{fig:lambda2p5}~(b), the applied flux is $A_{y,{\rm ext}}=0$, and the behavior of the gap components $\Delta_x$, $\Delta_y$ and of the current $J_y$  is similar to the low screening case below the transition, while Fig.~\ref{fig:lambda2p5}~(c) and (d) correspond to $A_{y,{\rm ext}} \simeq \mp 0.2 A_0$.
The behavior here is qualitatively different, and presents similarities with the curves observed above the transition in the low screening case: the gap $\Delta_y$ changes sign as a function of $x$ when going from one edge of the sample to the other, and the current $J_y$ is similar at the two edges of the system. 
Note that here, because of the strong screening, there is no region where the $\Delta_y$ component is zero. 
And the current density $J_y$ also has a marked feature in the region where the gap $\Delta_y$ is changing sign. 

These observations can be understood as due to the creation of a domain wall.
The middle (black) curves in Fig.~\ref{fig:lambda2p5}~(b) correspond to a system which is a $p_x + i p_y$ state on the whole $x$ axis, as the components $\Delta_x$ and $\Delta_y$ are both positive for all $x$ (see the inset, where the 
arrows represent schematically the direction of the edge
currents).
However, in Fig.~\ref{fig:lambda2p5}~(c) and (d), we see that $\Delta_y$ is changing sign for some $x$ value, meaning there is a domain wall between a $p_x + i p_y$ region and a $p_x- i p_y$ region. This is shown schematically in the insets, with the domain wall position shown as a dashed line.
This domain wall is manifest in the bumps visible in the current $J_{y}$ near the change of sign of $\Delta_y$. 
Fig.~\ref{fig:lambda2p5}~(c) and (d) correspond to a situation where a domain wall was created by the transition at high $|A_{y,{\rm ext}}|$, and remain present in the system when $A_{y,{\rm ext}}$, is slowly varied. 
The motion of the domain wall is illustrated
by the thin dotted curves of Fig.~\ref{fig:lambda2p5}c, which show $\Delta_y$ for
$A_{y,{\rm ext}} = -1.0$ and $A_{y,{\rm ext}} = 0.5$ (while the dashed red one is
for $A_{y,{\rm ext}} = -0.2$): one can see that the domain
wall, which correspond to the change of sign of $A_{y,{\rm ext}}$, is moving from left to right as $A_{y,{\rm ext}}$ is increased.


\subsection{Discussion}
\label{sec:Discussion}

The present results have been obtained on a simplified model where the system is at equilibrium, and the DC current is due to an applied flux.
However they could shed a new light on experimental results where current-voltage curves are measured.
Indeed, several experiments~\cite{Kambara08,Kambara10,Anwar13,Anwar17} have studied anomalous transport and critical current switching in Sr$_2$RuO$_4$ junctions. 
There, these anomalous switching have been shown to be related to the existence of chiral superconductivity, and the results have been discussed in terms of domain wall motion due to an applied DC current. 
Our results show that, in addition to domain wall motion, an applied DC current can \textit{create} a domain wall, and that the system can switch between different configurations with different numbers of domain walls (in our case 0 or 1), and which have different critical current values.
Fig.~\ref{fig:lambda2p5}~(b)-(d) show three different configurations which can exist for the same value of the applied flux, and which correspond to different values of the critical current.
The potential switching between these different states is reminiscent of the one observed experimentally when varying voltage.
A detailed comparison with the experiments would need to treat the nonequilibrium voltage states driven by the dynamics of the superconducting phase,\cite{Likharev79} which is beyond the scope of this work and is left for future study.

\section{Conclusions}
\label{sec:conclusion}

In this article, we have computed the response of a two-dimensional $p$-wave superconductor strip with a finite width under an external flux. 
We have performed quasi-classical calculations, where the Green function is obtained by solving the Eilenberger equation, using a Ricatti representation. 
The solution, obtained recursively,  gives access to the gap components $\Delta_x$, $\Delta_y$ and the current density $J_y$ as a function of the position.

As expected, applying a flux creates a DC current in the superconductor. 
We observe that above a given threshold for the applied flux, the system undergoes a transition, accompanied with the creation of a domain wall separating $p_x + i p_y$ and $p_x -i p_y$ regions. 
In the experimentally relevant case of strong screening, the system shows hysteresis: to a given value of the total current correspond several states which have different domain wall configurations. 

These results may be of importance for the understanding of  experiments measuring current-voltage characteristics of Sr$_2$RuO$_4$-Ru junctions\cite{Kambara08,Kambara10,Anwar13,Anwar17}, where critical current was shown to be related to chiral $p$-wave superconductivity. 
We think that domain wall creation by the applied DC current may be an important ingredient
to understand experimental results.
Extensions of this work could include the effect of surface roughness on the edge current\cite{Ashby09,Lederer14,Bakurskiy14,Suzuki16,Bakurskiy17}, the effect of impurities in the superconductor,\cite{Lu16}, and more realistic ring geometries.\cite{Jang11}
A study based on energetic arguments, in the spirit of Ref.~\citen{Vakaryuk}, studying
the coupling between domain wall currents and an applied vector potential also offers
promising perspectives.

\acknowledgments
We thank Shu-Ichiro Suzuki, Yasuhiro Asano, and Yuki Nagai for helpful discussion on the numerical method.
T.J. acknowledges support from CNRS PICS (7465, Japon 2017-2019).
T.K. acknowledges support by Grant-in-Aid for Scientific Research C (JP15K05124), and Grant-in-Aid for Scientific Research S (JP15K05153).

\appendix

\section{Details of the Numerical Calculations}
\label{appendix}

In numerical calculations, it is convenient to express all the quantities in a dimensionless form. 
We normalize the superconducting gaps, $\Delta_x(x)$ and $\Delta_y(x)$ by the zero-temperature bulk value $\Delta_0 \equiv \Delta_{\rm bulk}(T=0)$, and the length by the zero-temperature coherence length $\xi_0= \hbar v_F/\pi \Delta_0$.
We also normalize the vector potential $A_y(x)$, the magnetic field $B_z(x)$ and the current $J_y(x)$ 
by $\Delta_0/(e v_F)$, $\Phi_0/(2\sqrt{2}\pi \xi_0 \lambda_L(T))$, and $e v_F N(0) T_C$, respectively, where $\Phi_0 = h/2e$ is the flux quantum, $\lambda_L(T)=(\mu n_s e^2/m)^{-1/2}$ is the London penetration depth, $n_s(T)$ is the density of condensed electrons, and $m$ is the electron mass. 
Note that $T_c$ is here always the critical temperature in the absence of applied flux (the
variation of $T_c$ with the applied flux being very small).
In terms of dimensionless quantities, the Ricatti equations are written as
\begin{align}
\frac{da}{dx} &= \frac{1}{\pi \cos \theta_k} \left( \Delta \! -\! \Delta^* a^2 \! - \! 2 \left(\frac{\omega_n}{\Delta_0} 
      \! + \! i A_y \sin \theta_k \right) a \right). \\
\frac{db}{dx} &= \frac{1}{\pi \cos \theta_k} \left( -\Delta^* \! +\! \Delta \, b^2 \! + \! 2 \left(\frac{\omega_n}{\Delta_0} 
      \! + \! i A_y \sin \theta_k \right) b \right). 
\end{align}
The Maxwell equations, Eqs.~(\ref{eq:Bz})-(\ref{eq:Ay}), are written in terms of the dimensionless quantities as
\begin{align}
\frac{dB_z}{dx} &= - \frac{4 \sqrt{2}\xi_0}{\pi^2 e^{-\gamma}\lambda_L} J_y(x), \\
\frac{dA_y}{dx} &= \frac{\pi \xi_0}{2\sqrt{2}\lambda_L} B_z(x).
\end{align}
When integrating these equations, we impose the boundary conditions, $A_y(L/2) = A_{y,{\rm ext}}$ and $B_z(0) = \frac{2 \sqrt{2}\xi_0}{\pi^2 e^{-\gamma}\lambda_L} J_{tot}$,
$B_z(L) = -\frac{2 \sqrt{2}\xi_0}{\pi^2 e^{-\gamma}\lambda_L} J_{tot}$,
 where $A_{y,ext}$ is the external vector potential generated, and  $J_{tot}=\int_0^L dx J_y(x)$ is the total current.
Then, the magnetic field and the vector potential are obtained as
\begin{align}
B_z(x) &= - \frac{4 \sqrt{2}\xi_0}{\pi^2 e^{-\gamma}\lambda_L}  \left(\int_0^{x} \!\! dx' \ J_y(x') - \frac{1}{2} J_{tot} \right) , \label{eq:BzofJy} \\
A_y(x) &= - \int_x^{L/2} \!\! dx' \; \frac{\pi \xi_0}{2\sqrt{2}\lambda_L}B_z(x') + A_{y,{\rm ext}}.
\label{eq:AyofBz}
\end{align}

The iterative calculation is repeated until the variation of all the quantities between two successive steps is smaller than a given threshold.


\begin{thebibliography}{99}

\bibitem{Kallin16} C. Kallin and J. Berlinsky, Rep. Prog. Phys. {\bf 79}, 054502 (2016).

\bibitem{Maeno94} Y. Maeno, H. Hashimoto, K. Yoshida, S. Nishizaki, T. Fujita, J. G. Bednorz, and F. Lichtenberg, Nature {\bf 372}, 532 (1994).

\bibitem{Mackenzie03} A. P. Mackenzie and Y. Maeno, Rev. Mod. Phys. {\bf 75}, 657 (2003).

\bibitem{Maeno12} Y. Maeno, S. Kittaka, T. Nomura, S. Yonezawa, and K. Ishida, J. Phys. Soc. Jpn. {\bf 81}, 011009 (2012).

\bibitem{Matsumoto99} M. Matsumoto and M. Sigrist, J. Phys. Soc. Jpn. {\bf 68}, 994 (1999).

\bibitem{Furusaki01} A. Furusaki, M. Matsumoto, and M. Sigrist, Phys. Rev. B {\bf 64}, 054514 (2001).

%
\bibitem{Ivanov01} D. A. Ivanov, Phys. Rev. Lett. {\bf 86}, 268 (2001).

%
\bibitem{Nayak08} C. Nayak, S. H. Simon, A. Stern, M. Freedman, and S. Das Sarma, Rev. Mod. Phys. {\bf 80}, 1083 (2008).

%
\bibitem{Beenakker13} C. W. J. Beenakker, Annu. Rev. Condens. Matter Phys. {\bf 4}, 113 (2013).

%
\bibitem{DasSarma15} S. Das Sarma, M. Freedman and C. Nayak, npj Quantum Inf. {\bf 1}, 15001 (2015); arXiv:1501.02813.


\bibitem{Ishida98}
Ishida, K., H. Mukuda, Y. Kitaoka, K. Asayama, Z. Q. Mao, Y. Mori, and Y. Maeno, Nature (London) {\bf 396}, 658 (1998).

\bibitem{Luke98} G. M. Luke, Y. Fudamoto, K. M. Kojima, M. I. Larkin, J. Merrin, B. Nachumi, Y. J. Uemura, Y. Maeno, Z. Q. Mao, Y. Mori, H. Nakamura, and M. Sigrist, Nature (London) {\bf 394}, 558 (1998).


\bibitem{Xia06} J. Xia, Y. Maeno, P. T. Beyersdorf, M. M. Fejer, and A. Kapitulnik, Phys. Rev. Lett. {\bf 97}, 167002 (2006).

\bibitem{Kirtley07} J. R. Kirtley, C. Kallin, C. W. Hicks, E.-A. Kim, Y. Liu, K. A. Moler, Y. Maeno, and K. D. Nelson, Phys. Rev. B {\bf 76}, 014526 (2007).

\bibitem{Hicks10} C. W. Hicks, J. R. Kirtley, T. M. Lippman, N. C. Koshnick, M. E. Huber, Y. Maeno, W. M. Yuhasz, M. B. Maple, and K. A. Moler, Phys. Rev. B {\bf 81}, 214501 (2010).


\bibitem{Tada09} Y. Tada, N. Kawakami, and S. Fujimoto, New J. Phys. {\bf 11}, 055070 (2009).

\bibitem{Raghu10} S. Raghu, A. Kapitulnik, and S. A. Kivelson, Phys. Rev. Lett. {\bf 105}, 136401 (2010).

\bibitem{Huang14} W. Huang, E. Taylor, and C. Kallin, Phys. Rev. B {\bf 90}, 224519 (2014).

\bibitem{Komendova14} L. Komendov\'a and A. M. Black-Schaffer, Phys. Rev. Lett. {\bf 119}, 087001 (2017).

\bibitem{Taylor12}
E. Taylor and C. Kallin, Phys. Rev. Lett. {\bf 108} 157001 (2012).

\bibitem{Wysokinski12} K. Wysokinski, J. F. Annett, and B. L. Gy\"orffy, Phys. Rev. Lett. {\bf 108}, 077004 (2012).

\bibitem{Wysokinski13} M. Gradhand, K. K. Wysokinski, J. F. Annett, and B. L. Gy\"orffy, Phys. Rev. B {\bf 88}, 094504 (2013).

\bibitem{Imai12} Y. Imai, K. Wakabayashi, and M. Sigrist, Phys. Rev. B {\bf 85}, 174532 (2012).

\bibitem{Imai13} Y. Imai, K. Wakabayashi, and M. Sigrist, Phys. Rev. B {\bf 88}, 144503 (2013).

\bibitem{Scaffidi15} T. Scaffidi and S. H. Simon, Phys. Rev. Lett. {\bf 115}, 087003 (2015).

\bibitem{Kawai17} K. Kawai, K. Yada, Y. Tanaka, Y. Asano, A. A. Golubov, S. Kashiwaya, Phys. Rev. B {\bf 95}, 174518 (2017).

\bibitem{Zhang17a} L.-D Zhang, W. Huang, F. Yang, and H. Yao, arXiv:1710.00010.

\bibitem{Zhang17b} J.-L. Zhang, W. Huang, M. Sigrist, and D.-X. Yao, arXiv:1710.04519.




\bibitem{Kidwingira06} F. Kidwingira, J. D. Strand, D. J. Van Harlingen, and Y. Maeno, Science {\bf 314}, 1267 (2006).

\bibitem{Saitoh15} K. Saitoh, S. Kashiwaya, H. Kashiwaya, Y. Mawatari, Y. Asano, Y. Tanaka, and Y. Maeno, Phys. Rev. B {\bf 92}, 100504(R) (2015).	

\bibitem{Kambara08} H. Kambara, S. Kashiwaya, H. Yaguchi, Y. Asano, Y. Tanaka, and Y. Maeno, Phys. Rev. Lett. {\bf 101}, 267003 (2008).

\bibitem{Kambara10} H. Kambara, T. Matsumoto, H. Kashiwaya, S. Kashiwaya, H. Yaguchi, Y. Asano, Y. Tanaka, and Y. Maeno, J. Phys. Soc. Jpn. {\bf 79}, 074708 (2010).

\bibitem{Anwar13} M. S. Anwar, T. Nakamura, S. Yonezawa, M. Yakabe, R. Ishiguro, H. Takayanagi, and Y. Maeno, Sci. Rep. {\bf 3}, 2480 (2013).

\bibitem{Anwar17} M. S. Anwar, R. Ishiguro, T. Nakamura, M. Yakabe, S. Yonezawa, H. Takayanagi, and Y. Maeno, Phys. Rev. B {\bf 95}, 224509 (2017).


\bibitem{Vakaryuk}V. Vakaryuk, Phys. Rev. B \textbf{84}, 214524 (2011)

\bibitem{Schopohl98} N. Schopohl, arXiv:cond-mat/9804064 (unpublished).

\bibitem{Kopnin01} N. Kopnin, {\it Theory of Nonequilibrium Superconductivity} (Oxford University Press).

\bibitem{Nagai12} Y. Nagai, K. Tanaka, and N. Hayashi, Phys. Rev. B {\bf 86}, 094526 (2012).

\bibitem{Suzuki14} S.-I. Suzuki and Y. Asano, Phys. Rev. B {\bf 89}, 184508 (2014).

\bibitem{Likharev79} K. K. Likharev, Rev. Mod. Phys. {\bf 51}, 101 (1979).

\bibitem{Ashby09} P. E. C. Ashby and C. Kallin, Phys. Rev. B {\bf 79}, 224509 (2009).

\bibitem{Lederer14} S. Lederer, W. Huang, E. Taylor, S. Raghu, and C. Kallin, Phys. Rev. B {\bf 90}, 134521 (2014).

%
\bibitem{Bakurskiy14} S. V. Bakurskiy, A. A. Golubov, M. Y. Kupriyanov, K. Yada, and Y. Tanaka, Phys. Rev. B {\bf 90}, 064513 (2014).

\bibitem{Suzuki16} S.-I. Suzuki and Y. Asano, Phys. Rev. B {\bf 94}, 155302 (2016).

\bibitem{Bakurskiy17}S. V. Bakurskiy, N. V. Klenov, I. I. Soloviev, M. Yu Kupriyanov and A. A. Golubov, Supercond. Sci. Technol. {\bf 30} 044005 (2017).

\bibitem{Lu16} B. Lu, P. Burset, Y. Tanuma, A. A. Golubov, Y. Asano, and Y. Tanaka, Phys. Rev. B {\bf 94}, 014504 (2016).

\bibitem{Jang11} J. Jang, D. G. Ferguson, V. Vakaryuk, R. Budakian, S. B. Chung, P. M. Goldbart, Y. Maeno, Science {\bf 331}, 186 (2011).

\end{thebibliography}
\end{document}